\documentstyle[psfig]{mn}

\def\inv{\hbox{$^{-1}$}}

\begin{document}

\title[Correlation function of radio sources]{The correlation function of
radio sources}

\author[A.~J.~Loan, J.~V.~Wall and O.~Lahav]
  {A.~J.~Loan,$^1$
  J.~V.~Wall$^2$
  and O.~Lahav,$^1$\\
 $^1$ Institute of Astronomy, Madingley Road, Cambridge, CB3 0HA \\ 
 $^2$ Royal Greenwich Observatory, Madingley Road, Cambridge, CB3 0EZ \\  
}

\date{Received ; accepted , 1996}

\maketitle

\begin{abstract}
We investigate the large-scale clustering of radio sources in the Green Bank
and Parkes-MIT-NRAO 4.85\,GHz surveys by measuring the angular two-point
correlation function $w(\theta)$.  Excluding contaminated areas, the two
surveys together cover 70~per~cent of the whole sky.  We find both surveys to
be reasonably complete above 50\,mJy.  On the basis of previous studies, the
radio sources are galaxies and radio-loud quasars lying at redshifts up to $z
\sim 4$, with a median redshift $z \sim 1$.  This provides the opportunity to
probe large-scale structures in a volume far larger than that within the
reach of present optical and infrared surveys.  We detect a clustering signal
$w(\theta) \approx 0.01$ for $\theta=1\degr$.  By assuming an evolving
power-law spatial correlation function in comoving coordinates $\xi(r_c,z) =
( r_c / r_0 )^{-\gamma} \; (1+z)^{\gamma-(3+\epsilon)}$, where $\gamma=1.8$,
and the redshift distribution $N(z)$ of the radio galaxies, we constrain the
$r_0$--$\epsilon$ parameter space.  For `stable clustering' ($\epsilon=0$),
we find the correlation length $r_0 \approx 18\,h\inv$\,Mpc, larger than the
value for nearby normal galaxies and comparable to the cluster-cluster
correlation length.
\end{abstract} 

\section{Introduction}
 
\begin{figure*}
\centerline{\psfig{figure=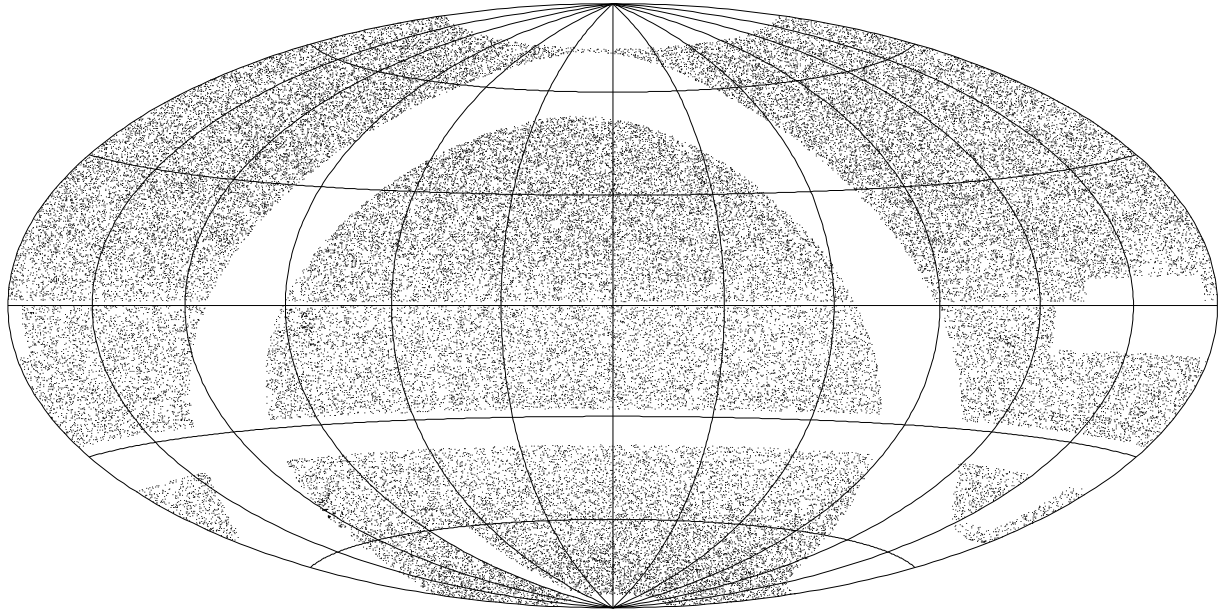,width=\textwidth,clip=}}
\caption{ The 87GB (northern hemisphere) and PMN (southern hemisphere) source
catalogues in Equatorial coordinates, Aitoff projection.  The Galactic Plane
($|b| < 10\degr$) and areas of poor coverage or confusion (see
Table~\protect\ref{cut-tab}) have been excised.  The varying source density
reflects the declination dependence of the flux limits of 87GB and PMN,
quantified in Figure~\protect\ref{surf-fig}.}
\label{slice-fig}
\end{figure*}

Surveys of galaxies selected by their optical or infrared emission reveal the
rich structure of the universe.  This clustering structure can be quantified
by using statistical techniques, such as the two-point correlation function
in two (e.g. APM: Maddox et al. 1990) or three (e.g. IRAS: Fisher et
al. 1993) dimensions.  However, these surveys only probe the local
universe:
for example, the median redshift of the APM survey is  $z \sim 0.1$.  In
contrast, radio galaxies and radio-loud quasars  can be detected over
significant cosmological distances, up to redshifts $z \sim 4$.  However,
such bright sources of radio emission represent a small fraction of all
galaxies, sampling the large volume within which they can be observed much
more sparsely than normal galaxies.

The prevailing consensus has long held that the distribution of radio
galaxies is almost isotropic, with any structures barely detectable.  Webster
(1976) studied the distribution of over 7500 sources from 3 separate radio
surveys at 148\,MHz, 408\,MHz and 1.4\,GHz, concluding that the data were
consistent with a uniform random distribution to within a few per~cent.
Shaver \& Pierre (1989) demonstrated that the distribution of
nearby ($z < 0.02$) radio galaxies was flattened towards the Supergalactic
Plane.  More recently, Peacock \& Nicholson (1991) analyzed a redshift survey
of 300 nearby ($z<0.1$) radio galaxies, measuring a power-law spatial
correlation function with a correlation length $r_0 = 11.0 \pm
1.2\,h\inv$\,Mpc, much larger than that for optically-selected galaxies ($r_0
\sim 5\,h\inv$\,Mpc -- Peebles 1980).  Benn \& Wall (1995) showed how the
apparent isotropy of radio sources may be used to estimate the largest scales
of cellular structure; further background and references are given therein.

The clustering of radio sources in the 4.85\,GHz Green Bank survey of the
northern hemisphere (87GB: Gregory \& Condon 1991) has been studied recently,
making use of the angular two-point correlation function $w(\theta)$
(Kooiman, Burns \& Klypin 1995; Sicotte 1995).  Both of these studies find
that there is clearly detectable clustering signal in the 87GB survey.  A
preliminary correlation function analysis of the first $\sim$15~per~cent of
the FIRST survey also finds significant clustering of radio sources (Cress
et al. 1996).

This paper presents a new study of the large-scale structures traced by the
radio sources in the 87GB survey.  We also present a correlation function
analysis of the Parkes-MIT-NRAO survey (PMN: Griffith \& Wright 1993).  The
PMN survey used much the same method as the 87GB survey to produce a
complementary 4.85\,GHz survey covering most of the southern sky.  With
appropriate selection of angular coverage and flux-density limit, the
two surveys prove to be quite well matched, covering most of the celestial
sphere.  Nevertheless, we analyze the surveys separately to avoid any
problems arising from small systematic differences in source detection
procedure and flux-density limit.

In Section~\ref{survey-sec} we examine the 87GB and PMN surveys to
demonstrate their suitability for correlation function analysis.  In
Section~\ref{2point-sec} we present the results of our analysis, while
Section~\ref{3point-sec} outlines the formalism that we use to estimate the
parameters of the correlation function in three dimensions.  In
Section~\ref{discuss-sec} we discuss the implications of our results and
compare them to those of other studies.

\section{The radio catalogues}

\label{survey-sec}

\subsection{The Green Bank survey (87GB)}

The 87GB source catalogue was compiled by Gregory \& Condon (1991) from
observations taken in 1987 October by Condon, Broderick \& Seielstad (1989)
using a seven-beam receiver attached to the (now defunct) NRAO 300~foot
(91\,m) radio telescope in Green Bank, West Virginia, USA.  The telescope was
tracked back and forth along the local meridian at $\pm 10\degr$\,min\inv,
the diurnal rotation of the earth eventually providing a Nyquist spacing of a
half beam-width between adjacent observations.  The data were reduced
automatically using a 5-$\sigma$ source-detection criterion to produce a
catalogue of 54,579 sources covering the declination band ($0\degr < \delta <
+75\degr$) for all right ascensions, except for four small confused or noisy
areas.  The 87GB survey has a total sky coverage of $\sim 6.0$\,steradians.

\subsection{The Parkes-MIT-NRAO survey (PMN)}

The PMN source catalogue was compiled from observations taken in 1990 June
and 1990 November with the ANRAO 64\,m radio telescope in Parkes, NSW,
Australia, covering the declination range $-87.5\degr < \delta < +10.0\degr$
for all right ascensions.  These observations followed the same observing
procedure using the same seven-beam receiver, and were reduced using an
algorithm similar to that used with 87GB.  However, a slightly different
source-detection criterion was adopted: 90~per~cent reliability --
equivalent
to $\sim$4.4-$\sigma$ -- rather than a 5-$\sigma$.

The source catalogue was split into 4 zones according to the elevation of the
telescope (see Table~1 of Wright et al. 1994): Southern, Tropical, Equatorial
and Zenith (respectively: Wright et al. 1994; Griffith et al. 1994; Griffith
et al. 1995; Wright et al. 1996).  However, because the Parkes telescope
tracks very slowly in the $10\degr$ wide region near the local zenith, the
data from the Zenith zone were observed using a different strategy to the
other three zones and is of a much lower quality.  For these reasons, the
Zenith zone is excluded from our analysis, leaving the other three zones
which together cover an total area of $\sim 6.4$\,steradians.

\subsection{Comparing the 87GB and PMN surveys}

\begin{table}
\caption{ Areas from the 87GB and PMN surveys included in and excluded
from our analysis.}
\begin{center}
\begin{tabular}{ll}
Included: & \\
87GB: & $75\degr > \delta > 1\degr$ \\
PMN:  & $0\degr > \delta > -28\degr$ \\
        & $-38\degr > \delta > -85\degr$ \\
Excluded: & \\
Galactic Plane:  & $|b| < 10\degr$ \\
Solar interference: & $6\degr > \delta > 1\degr$ and $|\alpha-205\degr| <20\degr$ \\
Extended sources:    & $\delta < -60\degr$ and $|\alpha-80\degr| < 10\degr$ \\
        & $\delta > -10\degr$ and $|\alpha-205\degr| < 30\degr$ \\
        & $-30\degr > \delta > -50\degr$ and $|\alpha-200\degr| < 10\degr$ \\
\end{tabular}
\end{center}
\label{cut-tab}
\end{table}

\begin{figure}
\centerline{\psfig{figure=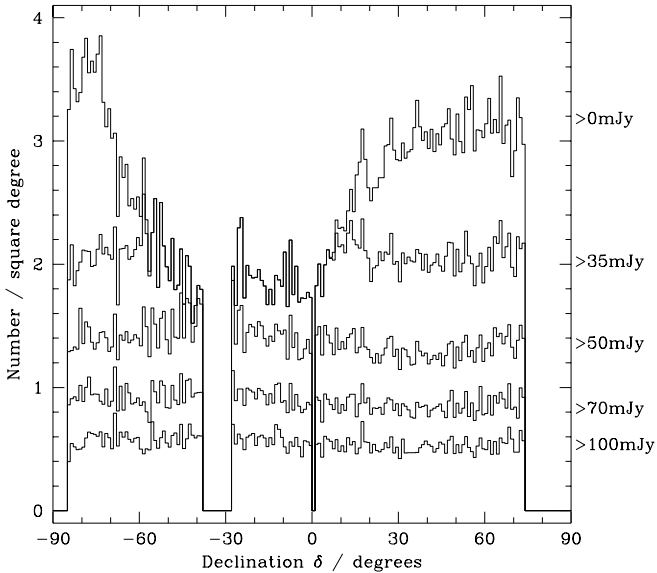,width=0.5\textwidth,clip=}}
\caption{ Histogram of the surface density of radio sources in the 87GB
($\delta>1\degr$) and PMN ($\delta<0\degr$) source catalogues as a function
of declination, for a range of flux-density limits (35, 50, 70 and
100\,mJy).}
\label{surf-fig}
\end{figure}

Before the 87GB and PMN source catalogues can be used for correlation
function analysis, we need to show that they are well matched, with no large
gradients caused by instrumental or other systematic effects.  We note that
real large-scale structures, such as the Supergalactic Plane, will introduce
inhomogeneities in these catalogues on large angular scales.  We will address
this issue elsewhere (Baleisis et al. 1996) but here assume that real
large-scale structures appear on small angular scales and use whole-sky
uniformity as our criterion for completeness.

The 87GB and PMN surveys overlap in the declination band $0\degr < \delta <
10\degr$ ($\approx 0.59$\,steradians).  Griffith et al. (1995) examined this
overlap region, finding agreement within the quoted errors between both the
published flux densities and the positions of the sources.  However, both
Griffith et al. (1995) and Baleisis (1996) report a worrying 2~per~cent
difference between the flux calibration of the 87GB and PMN catalogues, in
the sense that sources in the PMN survey are assigned a systematically lower
flux density than those in the 87GB survey.  For this reason, and to
highlight any discrepant results, we treat the surveys separately.  We note
that such a small calibration offset will not significantly affect the
results of our analysis on small scales, as we discuss below.

The flux-density limit of the 87GB survey varies with declination, increasing
from $\sim$25\,mJy at high declinations ($\delta > 60\degr$) to $\sim$40\,mJy
at the equator (Gregory \& Condon 1991), for two main reasons.  First, 
there is an
elevation dependence of the sensitivity of the 300\,foot Green Bank
telescope, and secondly, adjacent observing tracks become increasingly
further apart at lower declinations.  The Parkes telescope does not suffer
from the problem of decreased sensitivity at low elevations.  However, the
increased coverage due to adjacent scans lying closer together at
far-southern declinations ($\delta < -70\degr$) causes the flux-density 
limit to
decrease from $\sim$50\,mJy near the local zenith ($\delta\sim-33\degr$) to
$\sim$40\,mJy near the southern equatorial pole.

We use 87GB in the northern hemisphere ($\delta>1\degr$) because 87GB has a
flux limit which is lower than that for PMN over the region of overlap.  We
discard the region within $10\degr$ of the Galactic Plane to avoid
contamination by Galactic radio sources, and we also discard the area where
there is solar interference in the 87GB survey, together with three areas
with confusion from extended sources in the PMN catalogue.  The discarded
areas are shown in Table~\ref{cut-tab}.

The remaining sources, shown in Figure~\ref{slice-fig}, form the catalogue
which we use in our correlation analysis.  Figure~\ref{slice-fig} still shows
a strong variation in source density with declination due to changes in the
local flux-density limit.  To counteract this effect, we must impose a higher
flux limit over both surveys.

We plot the surface density of radio sources in the 87GB and PMN catalogues
as a function of declination in Figure~\ref{surf-fig} for flux-density limits
of 35, 50, 70 and 100\,mJy in order to find a safe lower flux-density limit,
which will provide a source catalogue with uniform sensitivity across the
whole area.  Both surveys are largely incomplete at the 35\,mJy flux
limits, with large density gradients caused by the sensitivity problems
discussed above.  The surface density of sources becomes increasingly more
uniform as the limit is raised from 50\,mJy, through 70\,mJy to 100\,mJy.
Above 50\,mJy, the surveys appear to be sufficiently evenly matched over the
whole sky to allow structure on small angular scales to be studied using
correlation techniques.  We will also calculate correlation functions at a
limit below 50\,mJy to demonstrate the effect of variable sensitivity on the
measured clustering.

There remains some suggestion of larger-scale structure present in the
combined 87GB/PMN source catalogue which we examine elsewhere (Baleisis et
al. 1996).

%%%%%%%%%%%%Section 3%%%%%%%%%%%%%%%%%%%%%%%%%%%%%%%%%%%%%%%%%%%%%%%%

\section{The angular correlation function}

\label{2point-sec}

\begin{figure*}
\centerline{\psfig{figure=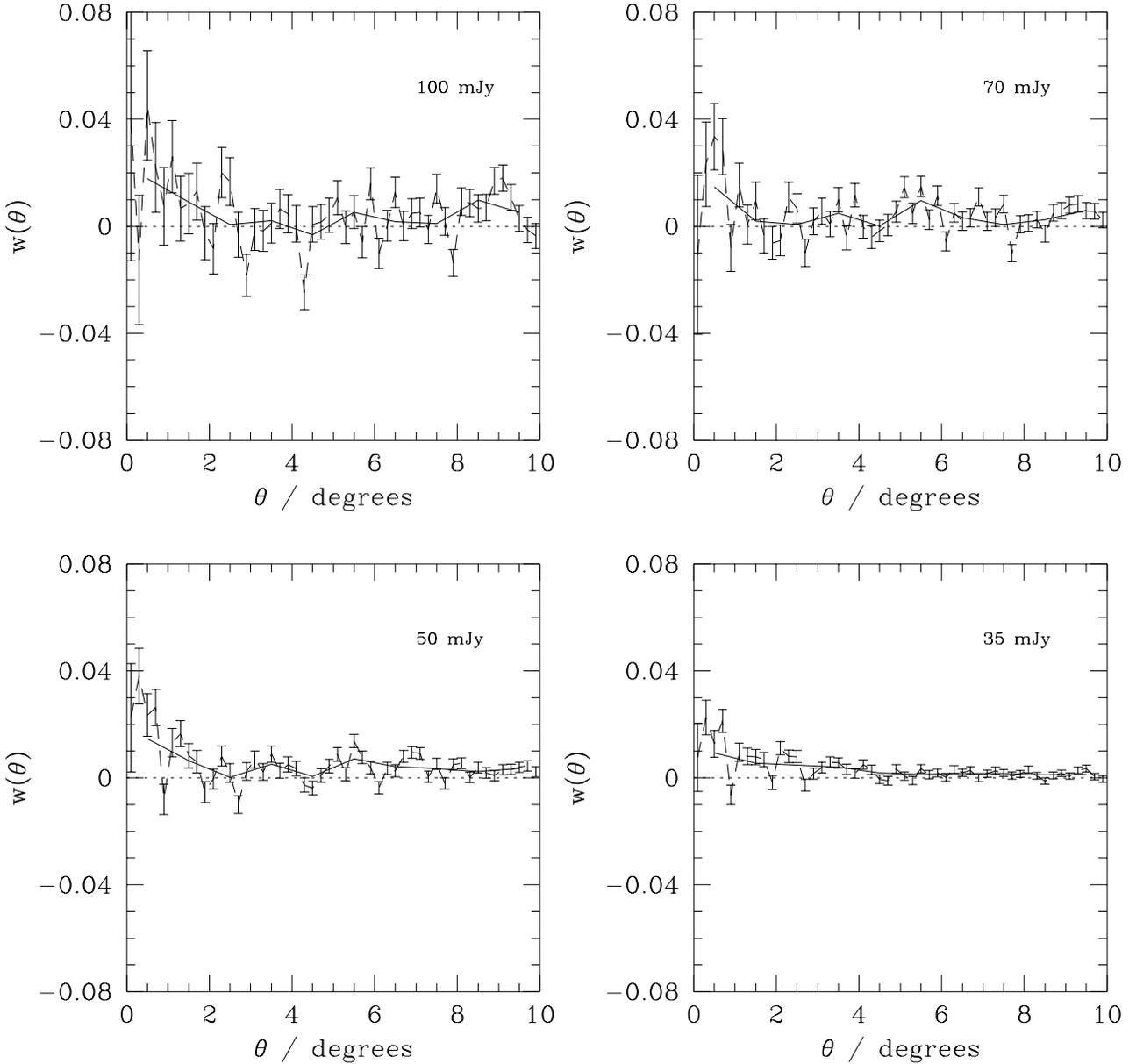,width=\textwidth,clip=}}
\caption{ Angular correlation function $w(\theta)$ for the 87GB
survey, with flux-density limits of 100\,mJy, 70\,mJy, 50\,mJy and
35\,mJy.  In each case the dashed line shows the data in $0.2\degr$
bins, and the solid line in $1\degr$ bins.  Poisson error bars are
shown on the line for $0.2\degr$ bins. The excess positive signal at
small angles ($\theta < 2\degr$) increases in amplitude at higher
flux-density limits.}
\label{w-gb-fig}
\end{figure*}

\begin{figure*}
\centerline{\psfig{figure=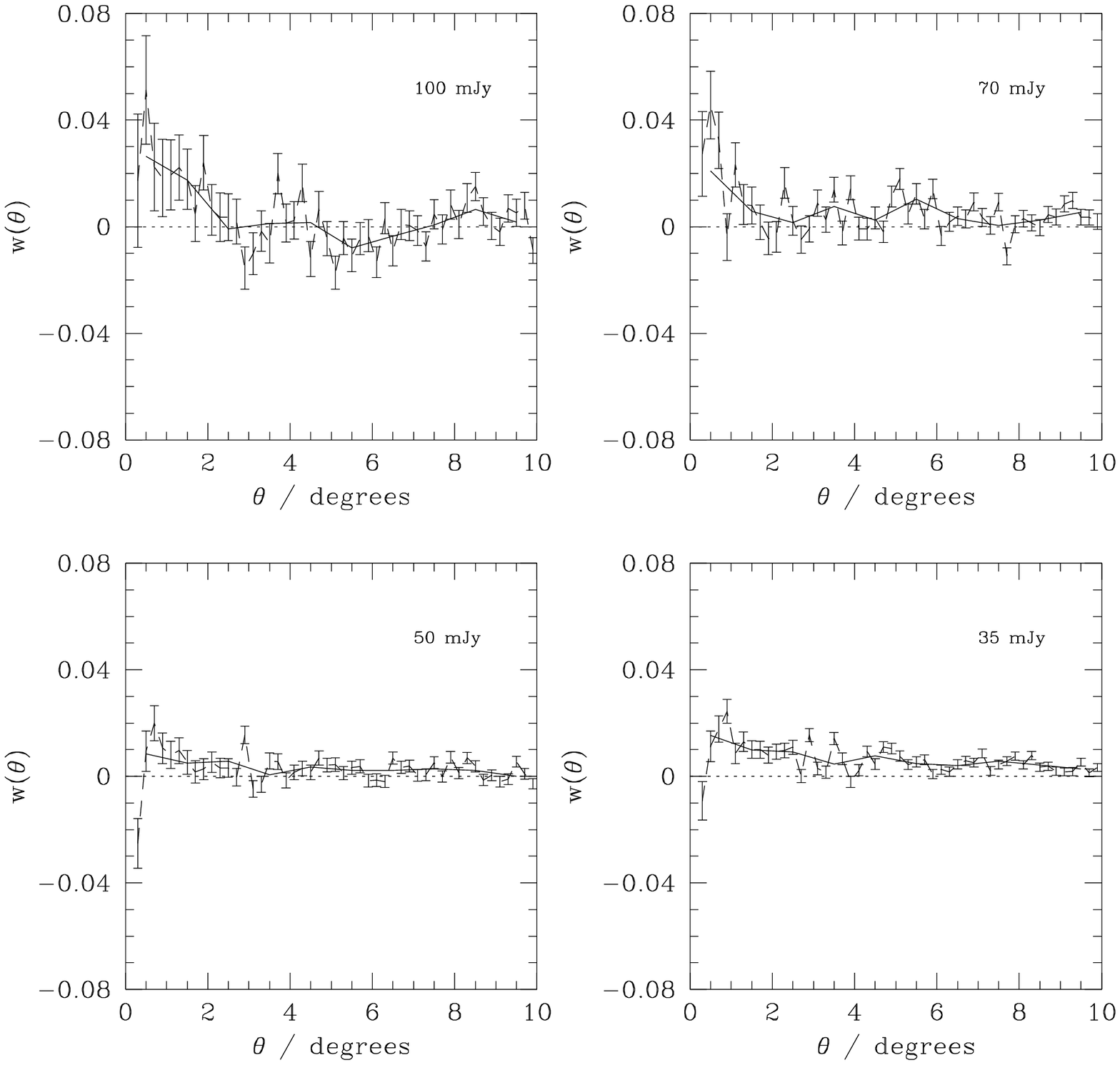,width=\textwidth,clip=}}
\caption{ Angular correlation function $w(\theta)$ for the PMN survey,
for flux-density limits of 100\,mJy, 70\,mJy, 50\,mJy and 35\,mJy.  In
each case the dashed line represents $0.2\degr$ bins and the solid
line $1\degr$ bins.  Poisson error bars are shown for the $0.2\degr$
binning. There is excess positive signal at small angles ($\theta <
2\degr$); the 35\,mJy plot shows the effect of the large density
gradients below a flux-density completeness limit.  }
\label{w-pmn-fig}
\end{figure*}

The two-point correlation function is a measure of how clustered the
sources are compared to a random Poisson distribution (e.g. Peebles
1980).  The angular two-point correlation function $w(\theta)$ gives
the excess probability of finding two sources in the solid angles
$\delta\Omega_1$ and $\delta\Omega_2$ separated by an angle $\theta$.
One way to estimate this function (Hamilton 1993) is to compare the
distribution of real galaxies to the distribution of points in a
random catalogue with the same boundaries, that is:
\begin{equation}
w(\theta) = \frac{DD \,.\, RR}{ (DR)^2 } - 1
\end{equation}
where $DD$, $RR$ and $DR$ are the numbers of data-data, random-random and
data-random pairs separated by the distance $\theta + \delta\theta$.  The
estimation of $RR$ and $DR$ requires a catalogue of `objects' scattered
uniformly over an area with the same selection properties as the data
catalogue (i.e. with the same angular boundaries).  The number of `objects'
in the random catalogue may be very much greater than that in the real
catalogues to reduce the errors in the estimates of $DR$ and $RR$.

We have calculated the angular two-point correlation functions for the 87GB
and PMN surveys separately, to highlight any discrepant results and also to
avoid problems if the catalogues prove to be mismatches.  The calculated
correlation functions for the 87GB and PMN catalogues are shown in
Figure~\ref{w-gb-fig} and Figure~\ref{w-pmn-fig}, respectively.  These
figures show an excess positive signal (i.e.  $w(\theta) > 0$) at small
angles ($\theta<2\degr$) with increasing amplitude at higher flux limits.
This increased amplitude arises from the `washing out' of structures by
foreground or background sources in surveys with a lower flux-density limit.
A shallower survey is expected to show a higher amplitude of the correlation
function since this `washing out' is less severe.  However, a survey with a
higher flux-density limits contains fewer sources, leading to larger errors
in the estimate of $w(\theta)$ which may conceal the greater expected
amplitude.

The correlation function for the 35\,mJy-limit PMN catalogue, fourth
panel of Figure~\ref{w-pmn-fig}, shows the effect of large-scale density
gradients in the survey noted from Figure~\ref{surf-fig}.  These gradients
are caused by choosing sources at a flux-density limit below that of the
catalogue completeness, producing a large error in the measured
correlation function.

\begin{figure}
\centerline{\psfig{figure=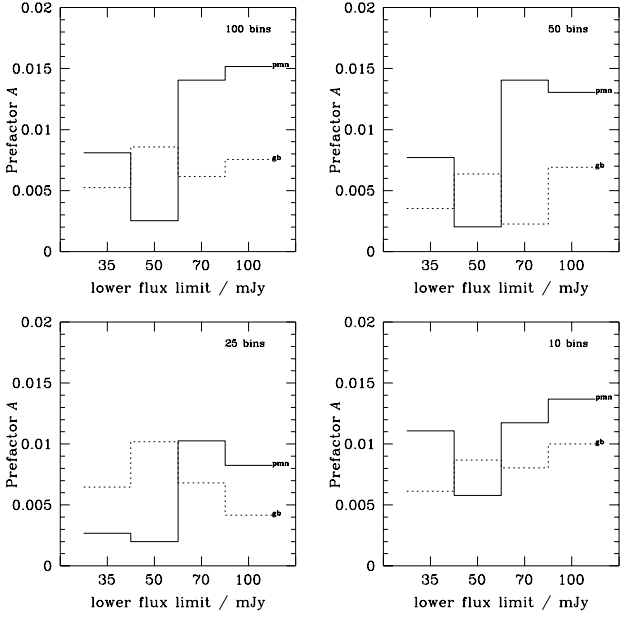,width=0.5\textwidth,clip=}}
\caption{ Estimated correlation function amplitude $A$ in the
expression $w(\theta) = A \theta^{1-\gamma}$ for a variety of
different binnings, for each chosen flux-density limit.  87GB is shown
with a dashed line, PMN with a solid line.  The lowest flux-density
limit, 35\,mJy, must be regarded as suspect due to incompleteness.
The remaining suggest a systematic trend with higher amplitude at higher
flux limit.}
\label{ampl-fig}
\end{figure}

We assume the usual power-law form for the angular correlation
function, such that
\begin{equation}
w(\theta) = A \theta^{1-\gamma} .
\end{equation}
If we assume $\gamma = 1.8$, as found for local galaxies (Peebles
1980), then only the amplitude $A$ of $w(\theta)$ remains to be
estimated from our measurements.  We have estimated this amplitude by
a simple least-squares procedure, by summing over the $i$ bins of our
calculated correlation function:
\begin{equation}
A_{\rm est} = \frac{\sum_{i} w_i \theta_i^{1-\gamma}}{\sum_{i} \theta_i^{2-2\gamma}} ,
\end{equation}
where $w_i$ is the observed value of $w(\theta)$ in the bin at angular
separation $\theta_i$.  Note that different bins of the observed
$w(\theta)$ are correlated.  Figure~\ref{ampl-fig} shows our estimates
for the amplitude of the correlation function for various binning
schemes and flux limits.  Our estimates indicate that the amplitude
$A$ has a value between 0.005 and 0.015.

%%%%%%%%%%%%%%SECTION 4%%%%%%%%%%%%%%%%%%%%%%%%%%%%%%%%%%%%%%%%%%%%%%%%%

\section{The spatial correlation function}
\label{3point-sec}

\subsection{The cosmological Limber's equation}

\label{limber-sec}

Limber's equation (Limber 1953) relates the spatial correlation function
$\xi(r,z)$ (in 3 dimensions) to the angular correlation function $w(\theta)$
(in 2 dimensions), given the selection function of the sample and a
cosmological model.  The cosmological Limber's equation can be written as
(e.g. Section 56 in Peebles 1980; Baugh \& Efstathiou 1993) :
\begin{equation}
w(\theta) = {\frac{ 2 \int_0^\infty \int_0^\infty x^4 F^{-2} \phi^2(x)
                \xi(r,z) dx du}{\left[ \int_0^\infty x^2 F^{-1}
                \phi(x) dx \right]^2 } }  ,
\label{limber-eq}
\end{equation}
where $x$ is a comoving coordinate and $F$ takes into account the different
possible world geometries (e.g. for $\Omega_0=1$, $\Lambda_0=0$, then $F=1$).
The selection function $\phi(x)$ satisfies the following relations for the
mean surface density in a survey of solid angle $\Omega_s$ :
\begin{equation}
{\cal N} = \int_0^\infty x^2 \phi(x) dx = \frac{1}{\Omega_s}
\int_0^\infty N(z) dz ,
\label{nofz-eq}
\end{equation} 
where $N(z) dz$ is the number of galaxies in the survey in the redshift shell
$(z, z+dz)$.

Hereafter we assume a flat universe with $\Omega_0 =1$, $F=1$, for which the
comoving distance is
\begin{equation}
x=\frac{2c}{H_0} [1 - (1+z)^{-1/2} ] ,
\label{xandz-eq}
\end{equation}
where $H_0 = 100\,h\inv$\,km\,sec\inv\,Mpc\inv\ is the Hubble expansion
parameter.  We also assume a power-law redshift-dependent 3-dimensional
correlation function:
\begin{equation}
 \xi(r,z) = \left(\frac{r}{r_0}\right)^{-\gamma} (1+z)^{-(3+\epsilon)} ,
\label{xi-pro-eq}
\end{equation}
where $r$ is a proper coordinate, $r_0$ is the correlation scale length at
redshift $z=0$, and $\epsilon$ describes the redshift evolution of the
spatial correlation function. In comoving coordinates, this can be written
as:
\begin{equation}
 \xi(r_c,z) = \left(\frac{r_c}{r_0}\right)^{-\gamma} (1+z)^{\gamma-(3+\epsilon)} ,
\label{xi-co-eq}
\end{equation}
where $r_c = r (1+z)$ is a comoving coordinate.  Specific values of
$\epsilon$ can be interpreted as follows: $\epsilon= 0$ corresponds to
constant clustering in proper coordinates, i.e. `stable clustering';
$\epsilon = \gamma-3 = -1.2$ implies constant clustering in comoving
coordinates; $\epsilon = \gamma-1 = 0.8$ implies growth of clustering under
linear theory (e.g. Peebles 1980; Treyer \& Lahav 1996).

For our purpose it is more convenient to express Limber's equation in term of
$N(z)$.  Using Equation~\ref{nofz-eq} and Equation~\ref{xandz-eq} we can
write:
\begin{equation}
\phi(x) = \frac{1}{\Omega_s} \frac{ H_0}{c} N(z) x^{-2} (1+z)^{+3/2} .
\end{equation}   
Substituting this into Equation~\ref{limber-eq} and assuming the small angle
approximation $r \approx (u^2/F^2 + x^2 \theta^2)^{1/2}/(1+z)$ we find that:
\begin{eqnarray}
w(\theta) &=& \left(\frac{r_0 H_0}{c}\right)^{\gamma} (2\theta)^{1-\gamma} H_{\gamma} \\ \nonumber 
&\times& \!\!\!\! \frac{ \int_0^\infty dz \left[ 1 - (1+z)^{-\frac{1}{2}} \right]^{1-\gamma}  N^2(z) \; (1+z)^{\gamma-\frac{3}{2} - \epsilon}} {\left[ \int_0^\infty dz \; N(z) \right]^2 } , 
\label{final-eq}
\end{eqnarray}
where $H_\gamma = \Gamma\left(\frac{1}{2}\right) \Gamma\left(
\frac{\gamma-1}{2} \right) / \Gamma\left( \frac{\gamma}{2} \right)$, with
$\Gamma$ the Gamma function.  Assuming $\gamma=1.8$, $H_{1.8} = 3.68$.

We emphasize that these equations are only valid for a power-law correlation
function of the form of Equation~\ref{xi-pro-eq} and $\Omega_0=1$ universe.
A more general expression is given in terms of a double integral and Fourier
transforms by Baugh \& Efstathiou (1993) and Treyer \& Lahav (1996).

Since $\gamma$ is assumed {\it a priori} to be $1.8$, the same as that for
normal galaxies, and $N(z)$ is determined from studies of the radio-source
population, the only free parameters to be deduced by comparison with the
observed $w(\theta)$ are the correlation scale length $r_0$ and the
clustering evolution index $\epsilon$.  Note that the amplitude of $N(z)$
does not affect Limber's equation; only the functional form of $N(z)$ is
important.

\subsection{The redshift distribution $N(z)$}

\label{nz-sec}

\begin{figure}
\centerline{\psfig{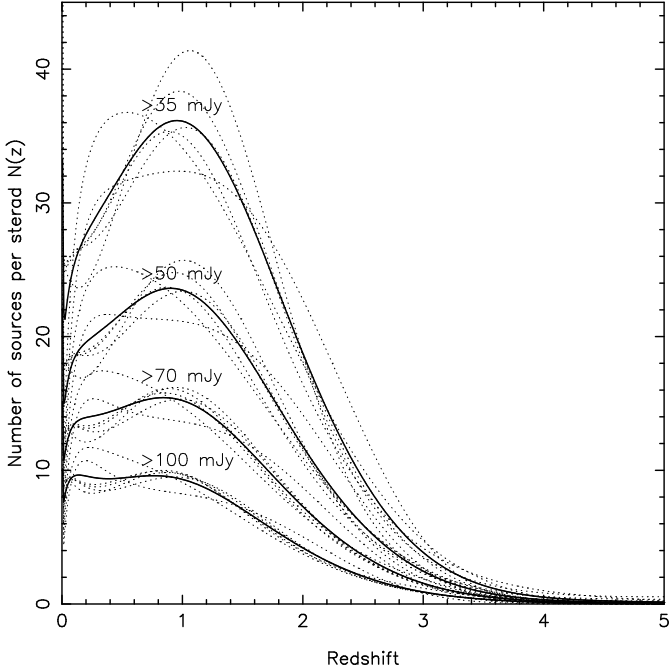}}
\caption{ Redshift distributions $N(z)$ of the radio-source population
selected at 4.85\,GHz, shown for each flux-density limit at which $w(\theta)$
has been calculated.  The solid curves represent the average of six models
taken from Dunlop \& Peacock (1990); see text.  The dotted curves for each
limit represent the 6 models from which the average was formed, indicating
the uncertainty at each flux-density level due to incomplete or
statistically-limited redshift data.  }
\label{nz-fig}
\end{figure}

The prediction of the angular two-point correlation function $w(\theta)$
from the spatial correlation function $\xi(r)$ requires knowledge of the
redshift distribution of the sources $N(z)$ complete to the designated
flux-density limit.  No complete sample of radio sources has been fully
identified at the flux-density level of 50\,mJy.  Despite this, Dunlop \&
Peacock (1990) have shown how a synthesis of the statistics of radio
sources (including source-counts together with incomplete identifications
and measured redshifts) may be used to determine the distribution of radio
sources with redshift, from which $N(z)$ may be derived at any flux
density.  In their analysis it was assumed that there are two independent
populations of radio sources, namely the flat-(radio)spectrum and
steep-spectrum objects.  The former are predominantly identified with QSOs,
the latter with radio galaxies; a more detailed overview of the
radio-source populations is given by Wall (1994). Treating the data for
flat- and steep-spectrum sources separately, Dunlop \& Peacock fit
polynomial expansions to describe the luminosity function and its
epoch-dependence for each population. They did so from seven different
starting points, the scatter amongst the seven models providing indication
of the statistical definition of the space density at any point over the
$P-z$ plane.

From these models we have calculated the $N(z)$ at 35, 50, 70 and 100\,mJy,
summing the space densities for the flat-and steep-spectrum populations.
At each flux-density limit, we then take the mean from the various models
as the the canonical value for $N(z)$.  (In this procedure we omitted model
5, which has a problem at the lower flux densities in that a dominating
spike appears in the space densities and $N(z)$ close to the redshift
cutoff which is specified as a starting condition for this particular
formulation.)  The $N(z)$ produced by this procedure is shown in
Figure~\ref{nz-fig}.  This figure indicates that the median redshift for
the radio sources in the 87GB/PMN survey is at $z\sim1$.
Figure~\ref{nz-fig} also shows that the average redshift decreases as the
lower flux limit is increased.

We note that a 2~per~cent flux mismatch between the catalogues has been
reported (Griffith et al. 1995; Baleisis 1996).  This mismatch would mean
that the flux limits given for the correlation functions estimated above are
slightly incorrect, and that we should use a different $N(z)$ corresponding
to the true flux in our Limber's inversion.  However, as we can see from
Figure~\ref{nz-fig}, the shape of $N(z)$ varies very little from one flux
limit to another, and so there will be little effect on the small angular
scales concerned.

\subsection{Estimating the spatial correlation function $\xi(r,z)$}

\label{epsilon-sec}

\begin{figure}
\centerline{\psfig{figure=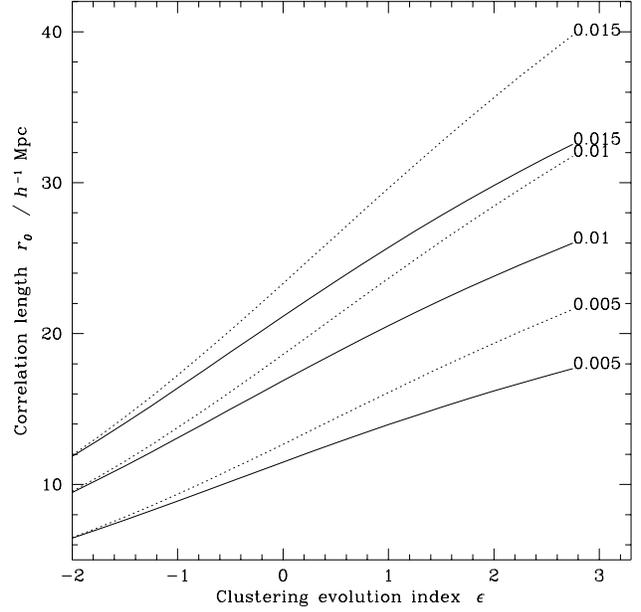,width=0.5\textwidth,clip=}}
\caption{ Predicted loci of constant $A$ in the correlation length $r_0$ and
clustering evolution index $\epsilon$ plane, for $S_{4.85} > 100$\,mJy (solid
lines) and $S_{4.85} > 50$\,mJy (dotted lines).  The values of $r_0$ and
$\epsilon$ which may be deduced from a single measurement of $A$ are strongly
correlated.}
\label{epsilon-fig}
\end{figure}

Armed with our estimate for the amplitude of the angular correlation function
from Section~\ref{2point-sec} ($A = 0.010 \pm 0.005$) and our estimate of
$N(z)$ from Section~\ref{nz-sec}, we can now compare this observed estimate
to the predictions from our model of the radio galaxy population for a range
of parameters, via equations 2 and 10.  Figure~\ref{epsilon-fig} shows lines
of constant $w(\theta)$ amplitude as a function of the spatial correlation
length $r_0$ and the evolution index $\epsilon$ at a flux-density limits of
50\,mJy and 100\,mJy.  There is a strong correlation between the values of
$r_0$ and $\epsilon$ which can be inferred from a single measurement of $A$.
Assuming a value for one parameter then provides an estimate for the other
parameter.  For example, consider $w(\theta=1\degr) = A = 0.01$ for a flux
limit of 50\,mJy.  If we also assume that clustering is constant in comoving
coordinates, i.e. $\epsilon = \gamma - 3 = -1.2 $, then $r_0 =
13\,h\inv$\,Mpc, while if clustering is `stable', i.e. $\epsilon = 0$, then
$r_0 = 18\,h\inv$\,Mpc.  The range of acceptable values of $A$ (0.005 to
0.015) translates into a uncertainty of a few Mpc in $r_0$.

\section{Discussion}
\label{discuss-sec}

Using two large-area 4.85\,GHz radio surveys with median redshift $z \sim 1$,
we have detected a correlation signal $w(\theta) \approx 0.01$ at angles
$\theta \sim 1\degr$ for radio sources brighter than 50\,mJy.  Adopting the
best estimate of redshift distribution for the sources, we find the
present-day correlation length to be in the range $13 < r_0 < 18\,h\inv$\,Mpc
for evolution index in the range $-1.2 < \epsilon < 0$, with uncertainty in
the measurements permitting errors as large as $\sim 5\,h\inv$\,Mpc.  The
values of $r_0$ and $\epsilon$ inferred from a measurement of $w(\theta)$ are
strongly correlated.

Previous studies by Kooiman et al. (1995) and Sicotte (1995) measured angular
correlation functions which were largely in agreement with each other and
with our results.  Kooiman et al. and Sicotte both attempted to use their
results to constrain the spatial correlation function of radio galaxies, but
reached contradictory results.  Sicotte found good agreement between his
model and his measured correlation function, with a best fit value of
$\epsilon = -1$ and $r_0 = 11\,h\inv$\,Mpc.  This result is in good agreement
our results, even though Sicotte used an earlier (Condon 1984) and
significantly different estimate of $N(z)$. (We repeated our analysis using
the Condon $N(z)$ and obtained remarkably similar results; there is little
sensitivity to the form of $N(z)$.)  This is also in accord with the value of
$r_0=11\,h\inv$\,Mpc measured by Peacock \& Nicholson (1991) from low
redshift ($z<0.1$) radio galaxies.

On the other hand, Kooiman et al. found that the amplitude of their model was
70 times smaller than the measured correlation function, assuming
$\gamma=1.8$, $\epsilon=0$ and $r_0 = 11\,h\inv$\,Mpc.  This is clearly
inconsistent with our results, and also with those of Sicotte.

Taken at face value, for a realistic range of the evolution index $\epsilon$,
we find a present-day correlation length $r_0$ which is far larger than that
for normal optical galaxies ($r_0 \sim 5\,h\inv$\,Mpc, Peebles 1980), IRAS
galaxies ($r_0 \sim 4\,h\inv$\,Mpc, Saunders, Rowan-Robinson \& Lawrence
1992), and QSOs ($r_0 \sim 6\,h\inv$\,Mpc for $\epsilon = -1.2$, Shanks \&
Boyle 1994).  However, our values of the correlation length for radio
galaxies is comparable to that for clusters of galaxies ($r_0 \sim
20\,h\inv$\,Mpc, Bahcall 1988).  This may reflect the preference of radio
galaxies at $z > 0.5$ to reside in high-density environments (Bahcall \&
Chokshi 1992; Yee \& Ellingson 1993 and references therein).

The uncertainty in our measurement of the correlation function could be
reduced by measuring the redshifts for at least a random subset of the
catalogues and cross-correlating this subset with the rest of the sample
(e.g. Saunders et al. 1992).  These redshifts would also improve our
knowledge of the redshift distribution of the radio galaxy population $N(z)$,
and indeed of the nature of the population (QSO - radio-galaxy - BL\,Lac
object) at this flux-density level.

Our analysis demonstrates that large-scale structure studies based on new
radio surveys nearing completion will provide decisive estimates of distant
structure and its evolution; and we have provided the appropriate
formalism. On-going surveys with the VLA -- FIRST (Becker, White \& Helfand
1995) and NVSS (Condon et al. 1996) -- will yield of order $10^6$ sources
over the sky, thus overcoming the limitations imposed by counting errors in
the estimated pair-counts from the current 4.85\,GHz surveys.  The first
correlation analysis of a subset of the FIRST survey (Cress et al. 1996)
indicates that the full survey will provide strong constraints on our models
of the radio source population. Furthermore, positional accuracy of these new
surveys will be such that direct spectroscopy at the catalogue radio
positions can be used to obtain redshifts.  Knowledge of the redshifts of even
a random subset of these catalogues will allow the use of
cross-correlation techniques to
improve the estimate of the correlation function in three-dimensions
(e.g. Saunders et al. 1992). 

\section*{Acknowledgments}

AJL acknowledges support from a PPARC studentship.  We thank Audra
Baleisis, Brian Boyle, Karl Fisher, Steve Maddox, and Jim Peebles for
helpful discussions, and John Peacock for generously making available
to us his software for calculating $N(z)$ from the Dunlop-Peacock
models.  The 87GB and PMN surveys are public data, generously made
available to the astronomical community by the original researchers.

\end{document}